\documentstyle[12pt,psfig]{article} 
\advance\textheight by\topskip 

\begin{document} 

\title{AN OVERVIEW OF THE SPOrt EXPERIMENT} 

\author{Michele Orsini, Ettore Carretti and Stefano Cortiglioni 
\and{\it C.N.R. - I.Te.S.R.E., Via Gobetti 101, 40129 Bologna - Italy}
\and{\it E-mail: surname@tesre.bo.cnr.it}} 
\date{} 
\maketitle 

\begin{abstract} 
The Sky Polarization Observatory is an experiment selected by
ESA for the Early Opportunity Phase onboard the International Space
Station.
SPOrt is the first payload specifically designed for polarization
measurements, it will provide near full sky maps of the sky polarized
emission at four microwave frequencies between 22 and 90 GHz.
Current design of SPOrt will be presented, together with an overview of
the scientific goals of the experiment.
\end{abstract}

\section {Introduction} 

The study of the polarized emission at microwave frequencies
is fundamental to understand the physical processes in
our Galaxy. Moreover, the galactic emission represents
a foreground noise in Cosmic Background Radiation (CBR)
anisotropy experiments.
Finally, the polarized
component of the CBR contains more information about
our universe than CBR anisotropies, in particular
concerning the nature of the primordial fluctuations and 
the re-ionization era.

The Sky Polarization Observatory (SPOrt) is a space experiment
devoted to measure the sky polarized 
emission in the microwave domain (20-90~GHz) 
with 7$^\circ$ beamwidth.
It is the first scientific payload specifically
designed to make a {\itshape clean} measurements
of the Q and U Stokes parameters.
SPOrt radiometers, in fact, are optimized for this purpose by
adopting:
\begin{description}
	\item{-} simple optics (corrugated feed horns)
to avoid additional spurious polarization from off-axis reflections;
	\item{-} low cross-polarization antenna system;
	\item{-} Q and U correlated outputs.
\end{description}

Main features of the SPOrt 
experiment are summarized in Table~\ref{main}, for
a detailed description of SPOrt technical issues
see \cite{cortiglio}.

During its 18 months lifetime, SPOrt will provide the 
first high frequency, high sensitivity 
($5 \div 10 \mu$K) polarization maps of the sky,
mapping the synchrotron emission at lower 
frequencies (20-30~GHz) and attempting the detection
of the polarized component of the CBR.

SPOrt is totally sponsored by the Italian Space Agency (ASI)
and it has been selected by the European Space Agency (ESA)
to fly on the International Space Station (ISS) 
in 2003.
\begin{table}
\caption{SPOrt characteristics}
\begin{center}
\label{main}
\begin{tabular}{ccccc}\hline
Frequencies  	&Bandwidth	& Angular 	&Instantaneous 		&Lifetime\\
(GHz)		&		& resolution	&sensitivity		&\\
\hline
22, 32, 60, 90	& 10\%		& $7^\circ$	&$1 mK s^\frac{1}{2}$	& 1.5 yrs\\
\hline
\end{tabular}
\end{center}
\end{table}

\section{The polarized emission of the sky}

Sky polarized emission comes both from 
non-cosmological (galactic and extragalactic)
 and from cosmological (CBR) sources.
In the following sections, the main
characteristics of both the galactic (``foreground'') and 
the CBR polarized emission are summarized, under the 
assumption that extragalactic 
polarized emission (coming from point sources) should be
considered negligible at SPOrt resolution.

\subsection{The galactic polarized emission}

The galatic emission arises from three different physical
processes: 
\begin{description}
	\item{-} synchrotron, which is produced by
relativistic electrons moving in the galactic
magnetic field;
	\item{-} bremsstrahlung (free-free), which 
comes from interactions between free electrons and
ions in a highly ionized medium;
        \item{-} dust emission, which has thermal origin.
\end{description}
So far data on polarized emission above 5~GHz are
lacking. Maps at 408, 465, 610, 820, 1411~MHz are
provided by \cite{spoelstra} and, 
more recently, low latitude galactic surveys
have been carried out by \cite{duncan} (at 2.4~GHz) and 
\cite{uyaniker1}, \cite{uyaniker2} (at 1.4~GHz).

This means that expected polarized emission at SPOrt
frequencies can be evaluated only by extrapolating 
low frequency data.

In general, predictions on both polarized 
and unpolarized sky emission are
different from author to author due to 
different assumptions on the normalization 
of the various foregrounds (see e.g. \cite{timbie}
and \cite{fabbri}).
There is, instead, a general agreement on foregrounds 
frequency behaviour. Synchrotron ($T_S$) and
free-free ($T_{F}$) brightness temperature 
are usually approximated by a power
law, while dust brightness temperature emission ($T_D$)
is modelled with a mixture of two greybodies
(cfr. \cite{kogut}):
\begin{equation} \nonumber
\begin{array}{rlrl} \nonumber
T_S &\propto \nu^S  & S &= - (2.6 \div 3.2)\\ \nonumber
T_F &\propto \nu^F  & F &\simeq - 2.15\\ \nonumber
T_D &\propto \nu^{D-2} \left[ B_\nu(20.4\mbox{K}) + 6.7 \cdot
   B_\nu(4.77\mbox{K}) \right] & D &\simeq 2
\end{array}
\end{equation}
While synchrotron and dust emission are intrinsicly
polarized, bremsstrahlung can be polarized only via
anisotropic Thomson scattering within optically thick HII regions;
expected polarization degrees on SPOrt angular scales
are $< 30 \%$ for synchrotron
and $\leq 10 \%$ for free-free and dust.

\begin{center}
\begin{figure}[t]
\hskip 2.5cm
\psfig{file=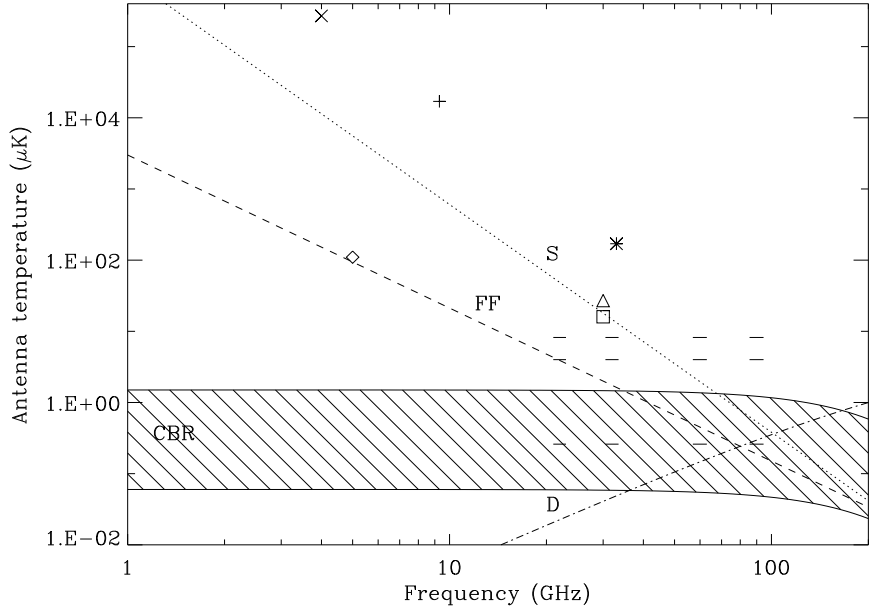,angle=90,width=12cm}
\caption{Expected sky polarized emission in the microwave domain: CBR 
(dashed strip) for two different values of normalization (5-0.2\%
of CBR temperature anisotropies i.e. maximum and minimum
$P(7^\circ)$ values in Figure~\ref{p_7deg}, respectively), 
synchrotron (dotted line), 
free-free (dashed line) and dust (dashed-dotted line); 
symbols show the upper limits on CBR polarization from previous 
experiments (see Table~\ref{others}); horizontal ticks show the 
expected pixel (min and max) and full sky sensitivities, respectively, for 
SPOrt at its four frequencies (see Table~\ref{sensitivities}).}
\label{sp_shape}
\end{figure}
\end{center}

Figure~\ref{sp_shape} shows the expected polarized
foregrounds. The following normalizations have 
been used: 
$T_S(30$~GHz$) = 18 \mu$K (dotted line), 
$T_F(30$~GHz$) = 2 \mu$K (dashed line), and 
$T_D(200$~GHz$) = 1  \mu$K (dash dotted line).
For a detailed discussion on foreground emissions see
\cite{fabbri}. However some remarks should be done:
\begin{description}
	\item{-} Maps at low frequencies show that 
the synchrotron power index varies with position. 
A frequency dependence of the spectral index is 
also expected: the break frequency of the synchrotron
spectrum depends on the break on the energy
distribution of cosmic rays and on the (spatially varying)
value of the galactic magnetic field.
	\item{-} Emission from spinning dust grains has been introduced
by \cite{draine} to explain correlations between 
$30 \div 40$~GHz galactic emission and DIRBE-FIRAS data (\cite{kogut2},
\cite{costa1}, \cite{costa2}). It would dominate over free-free and
synchrotron in the $10-60$~GHz frequency range and would be
intrinsicly polarized for $\nu < 40$~GHz (\cite{draine2}).
\end{description}

\subsection{The CBR Polarization}

The CBR is the most valuable witness of the early 
universe and CBR anisotropy investigations are a powerful 
tool to determine fundamental cosmological 
parameters such as the total density of the universe $\Omega_\circ$, 
the Hubble constant $H_\circ$, the baryon density $\Omega_b$ and 
the cosmological constant $\Lambda$ (\cite{white}).

Same informations could be provided by the CBR polarization 
which, in addition, represents 
the only way able to test the inflationary paradigm, 
to determine the nature (scalar or tensorial) of the primordial 
perturbations and to constraint the ionization history of 
the universe (see e.g. \cite{kam} and references therein).

The polarization of the CBR is generated by Thomson scattering of the 
anisotropic CBR radiation field in the cosmic medium, 
thus its intensity is only a fraction of the CBR temperature anisotropy;
depending on the epoch of the scattering, 
its spectrum shows the maximum power at different angular 
scales corresponding to the
particle horizon dimension at that epoch. 
For the Standard Cold Dark Matter (SCDM) 
model the peak occurs at $\leq 1^\circ$ angular scales, 
while for recombination models the maximum emission occurs at 
$5^\circ - 7^\circ$ angular scales.

\begin{center}
\begin{figure}
\hskip 3cm
\psfig{file=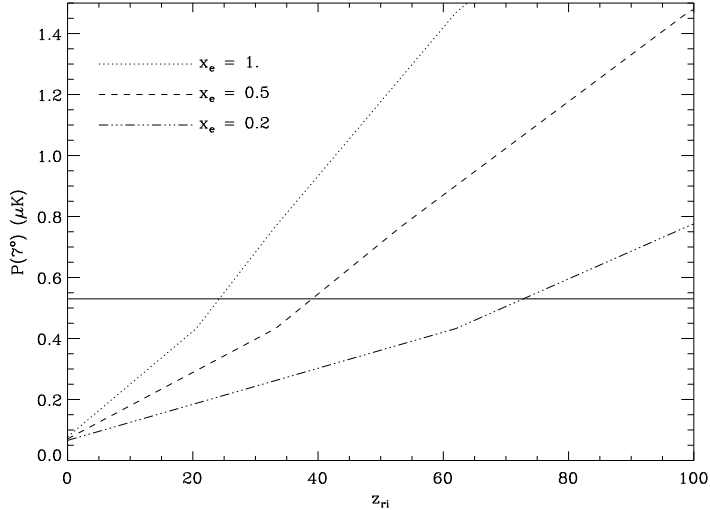,width=10cm,angle=90}
\caption{Expected CBR polarized emission (in terms of polarized
brightness temperature, $P = \sqrt{Q^2+U^2}$) smoothed to 7$^\circ$ 
calculated for $\Omega_b = 0.05$, $H_0 = 50$ vs reionization
redshift $z_{ri}$. Values are calculated for three different
ionization fractions ($x_e$). Solid horizontal line represents SPOrt full
sky sensitivity at $2\sigma$ level.}
\label{p_7deg}
\end{figure}
\end{center}

Expected polarized emission at $7^\circ$ angular scales for a SCDM and 
various recombination models are reported in Figure~\ref{p_7deg}. 
The minimum and the maximum values of $P(7^{\circ})$ 
represent the boundary of the dashed strip in Figure~\ref{sp_shape};
in this way foreground and CBR polarized emission can be compared easily.

None of the past and current CBR polarization experiments have 
led to a positive detection so far. 
Table~\ref{others} and Figure~\ref{sp_shape} report available upper 
limits only, in fact, the most stringent being $18 \mu$K on a 
$1^\circ.4$ scale near the North Celestial Pole (\cite{wollack}).
\begin{table}
\caption{Upper limits on CBR polarization from previous experiments.}
\begin{center}
\label{others}
\begin{tabular}{llllll}\hline
Frequency  	& Angular 		& Sky 					&Upper limit  		& Ref			&Fig. \ref{sp_shape}\\
(GHz)		& Resolution		& coverage				&			&			&\\
\hline
4.0	&15$^\circ$			&scattered				&300 mK			&\cite{penzias}		&X\\
100-600	&1$^\circ$.5 - 40$^\circ$	&GC 					&3-0.3 mK		&\cite{caderni}		&\\
9.3	&15$^\circ$			&$\delta$ = +40$^\circ$ 		&1.8 mK			&\cite{nanos}		&+\\
33	&15$^\circ$			&$\delta \in$ (-37$^\circ$, +63$^\circ$) &180 $\mu$K		&\cite{lubin}		&*\\
5.0	&18'' - 160''			&$\delta$ = +80$^\circ$ 		&4.2 mK - 120 $\mu$K	&\cite{partridge}	&diamond\\
26-36	&1$^\circ$.2			&NCP				 	&30 $\mu$K   		&\cite{wollack}		&triangle\\
26-36	&1$^\circ$.4			&NCP 					&18 $\mu$K  		&\cite{netterfield}	&square\\
\hline
\end{tabular}
\end{center}
\end{table}

\section{The SPOrt Correlation Receiver}

Correlation techniques are widely adopted in high
sensitivity measurements because of their capability
to reduce gain fluctuations.

Residual gain fluctuations are usually recovered, both for 
polarization and for total power experiments, using 
destriping techniques (see \cite{delabrouille}, 
\cite{wright}), which require a ``good'' radiometer 
stability within a single scan period.

Being not a free flyer, SPOrt has a scanning
period (i.e. the orbital period of the ISS $\sim$ 90 mins) larger
than other microwave space experiments (for instance COBE
and MAP have $\sim 1 \div 2$ min. scanning periods). In spite of this,
 the applicability of an efficient destriping
technique is based by the high stability
of the SPOrt correlation radiometers.

By correlating the linear ($E_x$ and $E_y$) and circular
($E_r$ and $E_l$) components of the incoming radiation
the following quantities can be obtained:
\begin{eqnarray} \nonumber
U  = 2 E_x  E_y  \cos{\epsilon} & &
   Q  = 2 E_r  E_l  \cos{\delta} = 2 E_r  E_l  \cos{2 \theta} \\
V  = 2 E_x  E_y  \sin{\epsilon} & &
   U  = 2 E_r  E_l  \sin{\delta} = 2 E_r  E_l  \sin{2 \theta} \\  \nonumber
\end{eqnarray}
where $\epsilon$ and $\delta$ are the phase difference
of the two linear and circular components respectively,
V is the Stokes parameter describing the circular
polarization, and $\theta$ is the polarization angle.

Thus, having to deal with linear components, 
a simultaneous correlated Q and U output cannot
be obtained. After a first measurement that gives U, in fact,  
the radiometer has to be rotated by $45^\circ$ in order
to have Q.
The correlation of circular components, instead,
provides a simultaneous measurement of the
Stokes parameters Q and U.

\begin{center}
\begin{figure}
\psfig{file=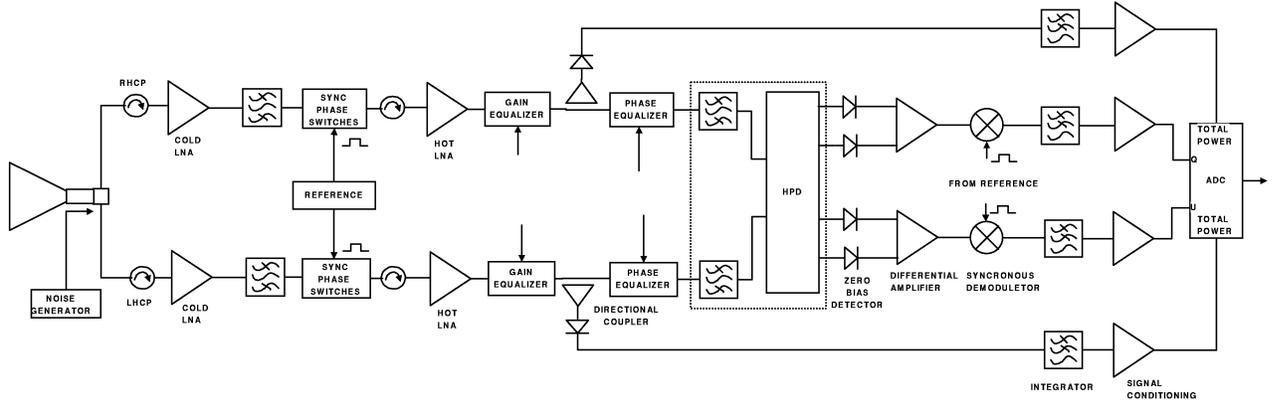,height=6cm,angle=-90}
\caption{Block diagram of one SPOrt polarimeter}
\label{block_diag}
\end{figure}
\end{center}

SPOrt adopts this second technique: a block diagram 
of a SPOrt radiometer is shown in
Figure~\ref{block_diag}. 

The antenna system provides left handed and right 
handed circular components to low noise radiometric
chains, which feed the correlation unit 
(i.e. Hybrid Phase Discriminator (HPD), zero bias diodes
and differential amplifiers).
The HPD outputs:
\begin{equation}
\left\{
\begin{array}{l}
\frac{1}{\sqrt{2}} \cdot \left( E_r-jE_l \right)\\
\frac{1}{\sqrt{2}} \cdot \left( E_l+jE_r \right)\\
\frac{1}{\sqrt{2}} \cdot \left( E_r-E_l \right)\\
\frac{1}{\sqrt{2}} \cdot \left( E_r+E_l \right)\\
\end{array}
\right.
\end{equation}
are then square law detected: 
\begin{eqnarray} \nonumber
V_1 & = & k \left[ \left( E_r^2 + E_l^2 \right) + 
   2 E_r E_l \cos{\delta} \right] \\  \nonumber
V_2 & = & k \left[ \left( E_r^2 + E_l^2 \right) - 
   2 E_r E_l \cos{\delta} \right] \\ \nonumber
V_3 & = & k \left[ \left( E_r^2 + E_l^2 \right) + 
   2 E_r E_l \sin{\delta} \right] \\ 
V_4 & = & k \left[ \left( E_r^2 + E_l^2 \right) - 
   2 E_r E_l \sin{\delta} \right] 
\end{eqnarray}
Finally, after differential amplification and integration, 
 Q and U are obtained: 
\begin{eqnarray} \nonumber
V_1 - V_2 & = & kE_r E_l \cos{\delta} \propto Q\\
V_3 - V_4 & = & kE_r E_l \sin{\delta} \propto U
\end{eqnarray}

\section{Expected performances}

\begin{center}
\begin{figure}
\hskip 4.5cm
\psfig{file=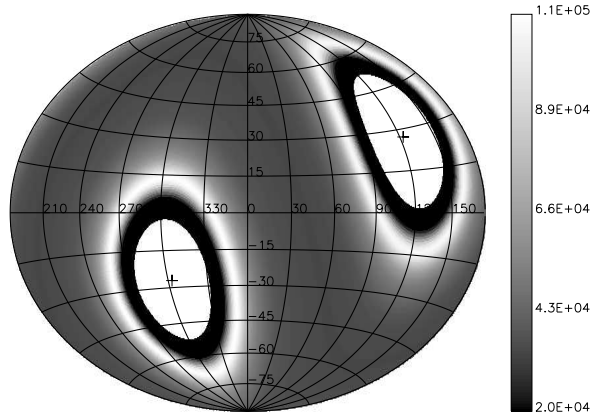,width=8cm}
\caption{SPOrt sky coverage after 18 months, the banner shows
the integration time in seconds after 1.5 yrs.}
\label{sport_coverage}
\end{figure}
\end{center}

A simulation of the SPOrt experiment sensitivity 
has been carried out (\cite{orsini}) 
using the radiometer equation:
\begin{equation}
\sigma_{f} = \frac{\sigma_{1s}}{\sqrt{\tau}}
\end{equation}
where $\tau$ is the total integration time and $\sigma_{1s}$
is the instantaneous radiometer sensitivity.
It showed how the scanning method of SPOrt makes the
sensitivity depending on the sky direction. 
Figure~\ref{sport_coverage} shows integration time across the 
sky after the expected 18 months experiment lifetime.
SPOrt final expected sensitivities are reported in 
Table~\ref{sensitivities} and in Figure~\ref{sp_shape},
 they are calculated assuming a 
50\% observing efficiency, i.e. rejecting half of the data;
``full sky'' sensitivity is calculated by averaging the 
signal over the whole sky.
\begin{table}
\caption{Expected sensitivities for SPOrt, 50\% observing efficiency 
has been assumed}
\begin{center}
\label{sensitivities}
\begin{tabular}{cccc}\hline
Max Pixel Sensitivity		& Min Pixel Sensitivity		& Full Sky Sensitivity		&Sky Coverage\\
\hline
 ($\mu$K)			&   ($\mu$K)			&   ($\mu$K)			&\\
4.0				& 8.0				& 0.26				& 82 \%\\
\hline
\end{tabular}
\end{center}
\end{table}

As it can be seen from Figure~\ref{sp_shape}, SPOrt frequencies
have been chosen to satisfy the two main experiment's aims:
\begin{description}
	\item{-} SPOrt lowest frequency channels (20-32~GHz) 
lie in a foreground dominated spectral window;
	\item{-} The highest frequency channels (60-90~GHz) could be
exploited to detect CBR polarized emission in a spectral window
were foreground emission has its minimum.
\end{description}

\section {Conclusions}

The SPOrt experiment is expected to measure the sky polarized 
emission in an unexplored frequency window with
expected sensitivity $\sim 50$ times better
than the best existing upper limit on CBR polarization.
SPOrt should reach these goals thanks to its wide 
frequency coverage that includes the ``cosmological window'' 
($\sim 60-100$~GHz) of minimum expected foreground emission.
Such a sensitivity, together with the extended sky ($>80$\%) and
frequency coverage should allow to put new and quite severe constraints
on the CBR polarization at degrees angular scales.
Finally, it would be pointed out that the very simple SPOrt 
layout configuration has been mainly
imposed by the need to match the ISS environment, but it
represents anyway a new approach to very sensitive 
polarization measurements.

\vskip 0.5cm
\noindent
{\bf Acknowledgements}
Authors acknowledge ESA for the encouragement and partial 
financial support of the SPOrt project, as well as ASI for the 
full approval and funding of the SPOrt Program.
A special thank is for all the SPOrt collaboration.
Figure~\ref{p_7deg} has been produced by using the 
CMBFast (\cite{cmbfast})
software and Healpix package (\cite{healpix}).

{} 


\begin{thebibliography}{} 

\bibitem{cortiglio}
S. Cortiglioni, S. Cecchini, E. Carretti, M. Orsini, R. Fabbri, G. Boella, 
G. Sironi, M. Gervasi, M. Zannoni, J. Monari, A. Orfei, R. Tascone, 
U. Pisani, K.W. Ng, L. Nicastro, L. Popa, I.A. Strukov, M.V. Sazhin
in proceedings of the International Conference on 3K Cosmology
EC-TMR Conference, AIP Conference Proceedings, 476, pp.194-203, 1998
 - astro-ph/9901362.

\bibitem{spoelstra}
W.N. Brouw, T.A.Th. Spoelstra, Astronomy and Astrophysics 
Supplement, 26, pp. 129-146, 1976.

\bibitem{duncan}
A.R. Duncan, R.F. Haynes, K.L. Jones, R.T. Stewart, MNRAS, 291, 
pp. 279-295, 1997.

\bibitem{uyaniker1}
B. Uyaniker, E. Fuerst, W. Reich, , P. Reich, R. Wielebinski, 
Astronomy and Astrophysics Supplement, 132, pp. 401-411, 1998.

\bibitem{uyaniker2}
B. Uyaniker, E. Fuerst, W. Reich, , P. Reich, R. Wielebinski, 
Astronomy and Astrophysics Supplement, 138, pp. 31-45, 1999.

\bibitem{timbie}
B. Keating, P. Timbie, A. Polnarev, J. Steinberger, 
The Astrophysical Journal, 495, pp. 580, 1998.

\bibitem{fabbri}
R. Fabbri, S. Cortiglioni, S. Cecchini, M. Orsini, E. Carretti, G. Boella, 
G. Sironi, J. Monari, A. Orfei, R. Tascone, U. Pisani, K.W. Ng, 
L. Nicastro, L. Popa, I.A. Strukov, M.V. Sazhin
in proceedings of the International Conference on 3K Cosmology 
EC-TMR Conference, AIP Conference Proceedings, 476, pp.194-203, 1998
 -- astro-ph/9901363.

\bibitem{kogut}
A. Kogut, A.J. Banday, C.L. Bennett, K.M. Gorski, G. Hinshaw, 
W.T. Reach, The Astrophysical Journal, 460, pp. 1-9, 1996.

\bibitem{draine}
B.T. Draine, A. Lazarian, The Astrophysical Journal,
494, pp. L19-L22, 1998.

\bibitem{kogut2}
A. Kogut,  astro-ph/9902307, 1999.

\bibitem{costa1}
A. De Oliveira-Costa, M. Tegmark, C.M. Gutierrez, A.W. Jones, R.D. Davies,
A.N. Lasenby, R. Rebolo, R.A. Watson, astro-ph/9904296, 1999.

\bibitem{costa2}
A. De Oliveira-Costa, M. Tegmark, L.A. Page, S.P. Boughn,
The Astrophysical Journal, 509, pp. L9-L12, 1998.

\bibitem{draine2}
B.T. Draine, A. Lazarian, astro-ph/9902356, 1999.

\bibitem{white}  
M. White, D. Scott, J. Silk, Annual Review of Astronomy and 
Astrophysics, 32, pp. 319-370, 1994.

\bibitem{kam}
M. Kamionkowski, A. Kosowsky, A. Stebbins, Phys. Rev. D, 55,
7368, 1997.

\bibitem{lubin} 
P.M. Lubin, G.F. Smoot, The Astrophysical Journal, 245, pp. 1-17,
1981.

\bibitem{penzias}  
A.A. Penzias, R.W. Wilson, The Astrophysical Journal,
142, pp. 419-421, 1965.

\bibitem{caderni}  
N. Caderni, R. Fabbri, F. Melchiorri, V. Natale, 
Phys. Rev. D, 17, pp. 1901-1907, 1978.

\bibitem{nanos}  
G.P. Nanos, The Astrophysical Journal, 232, 
pp. 341-347, 1979.

\bibitem{partridge}  
R.B. Partridge, J. Nowakowski, H.M. Martin, 
Nature, 331, pp. 146-147, 1988.

\bibitem{wollack}  
E.J. Wollack, N.C. Jarosik, C.B. Netterfield, L. Page,
D. Wilkinson,  The Astrophysical Journal, 419, 
pp. L49-L52, 1993.

\bibitem{netterfield}  
C.B. Netterfield, N. Jarosik, L. Page, D. Wilkinson, E.J.
Wollack, The Astrophysical Journal, 445, pp. L69-L72, 1995.

\bibitem{delabrouille}
J. Delabrouille, Astronomy and Astrophysics Supplement, 
127, pp. 555-567, 1998.

\bibitem{wright}
E.L. Wright, astro-ph/9612006.

\bibitem{orsini}
M. Orsini, L. Nicastro, E. Carretti, S. Cortiglioni
I.Te.S.R.E. C.N.R. Internal Report, \#233/99, 1999.

\bibitem{cmbfast}
See CMBFAST Website:

http:\-//\-www.\-\-sns.\-\-ias.\-\-edu\-/$\sim$matiasz\-/CMBFAST\-/cmbfast.html

\bibitem{healpix}
See the Healpix home page:

http:\-//\-www.\-\-tac.\-\-dk\-/$\sim$healpix

\end{thebibliography}
\end{document}